\documentclass[12pt]{article}
\usepackage{epsfig}
\usepackage{amsmath}
\usepackage{graphicx}
\usepackage{latexsym}

\begin{document}

\begin{center}
{\Large \bf The influence of quantum field fluctuations \\ on
chaotic dynamics of Yang-Mills system II. \\\vspace{0.2cm} The role
of the
centrifugal term}\\
\vspace{0.5cm}
{\large {\bf V.I. Kuvshinov, A.V. Kuzmin}}\\
\textsf{Joint Institute for Power and Nuclear Research},\\
{\it   acad. A.K. Krasina 99, Minsk,  220109, Belarus} \\
{\rm  E-mail: V.Kuvshinov@sosny.bas-net.by,\\
avkuzmin@sosny.bas-net.by}\\
\vspace{0.5cm}
{\large {\bf V.A. Piatrou}}\\
\textsf{Belarusian State University},\\
 {\it 2 Nezalejnasci av., Minsk, 220050, Belarus} \\
{\rm  E-mail:     PiatrouVadzim@tut.by}
\end{center}

\begin{abstract}
We have considered $SU(2)\bigotimes U(1)$ gauge field theory
describing electroweak interactions. We have demonstrated that
centrifugal term in model Hamiltonian increases the region of
regular dynamics of Yang-Mills and Higgs fields system at low
densities of energy. Also we have found analytically the approximate
relation for critical density of energy of the order to chaos
transition on centrifugal constant. It is necessary to note that
mentioned increase of the region of regular dynamics has linear
dependance on the value of the centrifugal constant.
\end{abstract}

A steady interest to chaos in gauge field theories \cite{BookGFT}
is connected with the fact that all four fundamental particle
interactions have chaotic solutions \cite{4}. There are a lot of
footprints of chaos in HEP \cite{Kawabe, N}, nuclear physics
(energy spacing distributions) \cite{Nuclear, Bunakov}, quantum
mechanics \cite{QM}.

Much attention has been paid in the last decade to chaos in
quantum field theory. Non-abelian Yang-Mills gauge fields were
investigated without spontaneous symmetry breaking. It was
analytically and numerically shown that classical Yang-Mills
theories are inherently chaotic ones\cite{MST,Savvidy}. The
further research has shown for spatially homogeneous field
configurations \cite{SHS} that spontaneous symmetry breakdown
leads to appearance of order-chaos transition with rise of density
of energy of classical gauge fields \cite{regular,SavvidyNucl},
whereas dynamics of gauge fields in the absence of spontaneous
symmetry breakdown is chaotic at any density of energy
\cite{Savvidy}.

 In the work \cite {Previous} it was shown that the "switching on" of
quantum fluctuations of vector gauge fields leads to ordering at low
densities of energy, order-to-chaos transition with the rise of
density of energy of gauge fields. Also it was noted that if the
ratio of the coupling constants of Yang-Mills and Higgs fields is
larger than some critical value then quantum corrections do not
affect the chaotic dynamics of gauge and Higgs fields.

In this paper we investigate the influence of the centrifugal term
in model Hamiltonian on chaotic dynamics of Yang-Mills and Higgs
gauge fields. We demonstrate numerically and analytically that
centrifugal term increases the energy region of regular dynamics
of this fields.

Consider $SU(2)\bigotimes U(1)$ gauge field theory which is
describing electroweak interactions with real massless scalar field
$\rho$ with the Lagrangian
\begin{align}\label{Lagrangian}
 L=&-\frac{1}{4}G_{\mu \nu }^{a}G^{a}{}^{\mu \nu }-\frac{1}{4}H_{\mu \nu
}H^{\mu \nu }+\frac{1}{8}g^{2}\rho ^{2}\left( W_{1}^{2}+W_{2}^{2}+\frac{W_{3}^{2}%
}{\cos ^{2}{\theta _{w}}}\right) \notag
\\
\qquad &+\frac{1}{2}\partial _{\mu }\rho \partial ^{\mu }\rho -
\frac{1}{4!}\lambda \rho ^{4}, \quad a=1,2,3, \quad \mu=1,2,3,4.
\end{align}

We used the following denotations
\begin{equation}
G^{a}_{\mu \nu}=\partial_{\mu}W^{a}_{\nu} -
\partial_{\nu}W^{a}_{\mu} +
g\varepsilon^{abc}W^{b}_{\mu}W^{c}_{\nu};
\end{equation}
\begin{equation}
H_{\mu \nu} = \partial_{\mu}A_{\nu} -
\partial_{\nu}A_{\mu}.
\end{equation}
Here  $W_{\mu }^{1}$, $W_{\mu }^{2}$ describe $W^{\pm}$-bosons and
$W^{3}_{\mu}$ - neutral Z-boson, $A_{\mu }$ corresponds
electro-magnetic field, $g$ denotes a self-coupling constant of
non-abelian gauge fields, $\lambda $ - self-coupling constant of
scalar field,  $\theta _{w}$ is Weinberg angle.

To study the dynamics of classical gauge fields from the viewpoint
of chaos we simplify the problem using spatially homogeneous
solutions \cite{SHS}. So we will investigate fields of the
following form
\begin{equation}\label{simplification}
W_{i}^{a}=e_{i}^{a}q_{a}\left( \tau \right),\quad a=1,2; \quad
i=1,2,3; \quad
\overrightarrow{e^{a}}=\left(e_{1}^{a},e_{2}^{a},e_{3}^{a}\right)
,\quad \left(\overrightarrow{e^{a}}\right)^{2}=1.
\end{equation}
Here $\overrightarrow{e^{a}}$ are constant unit vectors which is
describing the linear polarization of the gauge fields
$\overrightarrow{W^{a}}$. Also we use simplification:
\begin{equation}\label{simplification2}
\overrightarrow{e^{1}}\overrightarrow{e^{2}}=1,
\end{equation}
which  means that the fields of $W^+$ and $W^-$ gauge bosons have
the same linear polarization. Other classical gauge fields for
simplicity are put to be equal to zero

It was shown that the dynamics of classical gauge fields in the
classical vacuum of scalar field $\rho =0$ is chaotic at any
densities of energy \cite{Savvidy}. Situation qualitatively changes
if we take into account quantum fluctuations of vector fields. It is
caused by the known fact that the state $<\rho > = 0$ is not a
vacuum state in this case. Here $<\rho>$ denotes vacuum quantum
expectation value of scalar field related with its classical vacuum
value $\rho$ as follows $<\rho
>= \rho + quantum \quad corrections$. To find a true vacuum of
scalar field in this case we use the method of the effective
potential \cite{Coleman}.

One loop  effective potential generated by the Lagrangian
(\ref{Lagrangian}) has the form (see also \cite{Previous},
\cite{Huang})
\begin{align}\label{eff}
 U\left( <\rho> \right)
= &\frac{1}{4!}\lambda <\rho >^{4} +
\frac{3g^{4}<\rho > ^{4}}{128\pi ^{2}}\left( -\frac{1}{2}+\ln {\frac{%
g^{2}<\rho > ^{2}}{2\mu ^{2}}}\right)  \notag \\
\qquad &+\frac{3g^{4}<\rho > ^{4}}{256\pi ^{2}\cos ^{4}{\theta _{w}}}\left( -\frac{1}{2}%
+\ln {\frac{g^{2}<\rho >^{2}}{2\mu ^{2}\cos ^{2}{\theta
_{w}}}}\right).
\end{align}
Where $\mu^{2}$ is a renormalization constant. Here we took into
account contributions of all Feynman diagrams with one loop of any
($W_{1}, W_{2}, W_{3}$ or $A$) gauge field and external lines of
Higgs field. This potential leads to spontaneous symmetry breaking
and non-zero vacuum expectation value of scalar $\rho-$field
appears. Classical vacuum of scalar field is $\rho =0$, but it is
not so in quantum case, because of Coleman-Weinberg effect
\cite{Coleman}. In simplifications (\ref{simplification}) we put
classical gauge fields $\vec{W^{3}}$ and $\vec{A}$ to be equal to
zero,  but their quantum fluctuations are included and give
contribution to the potential (\ref{eff}).

Hamiltonian describing dynamics of the model field system has the
following form

\begin{equation}\label{NewH}
  H = \frac{1}{2} (p_r^2 + p^2+\frac{M^2}{ r^2}) + \frac{1}{8} g^{2} <\rho>^2 r^2 +
  U(<\rho>),
\end{equation}
where we make the substitution $q_1 = r \cos{\varphi}$ and $q_2 = r
\sin{\varphi}$. It is clarify that $\varphi$ is a cyclical variable
and therefore its conjugate momentum $p_\varphi = r^2 \dot{\varphi}
= M$ is a constant of motion. Also we used denotations $p_r^2 =
\dot{q}_1^2 + \dot{q}_2^2$, $\quad r^2 = q_1^2 + q_2^2$. Here $p_r$
is a momentum of gauge fields  and $p$ is a momentum of Higgs field.

In contrast to article \cite{Previous} we didn't neglected here by
the term ${M^2}/{r^2}$ and investigated the influence of the
centrifugal term ${M^2}/{r^2}$ on the dynamics of the model field
system.

\begin{figure}[t!]
  \centering
  \leavevmode
\includegraphics[height=5.5cm]{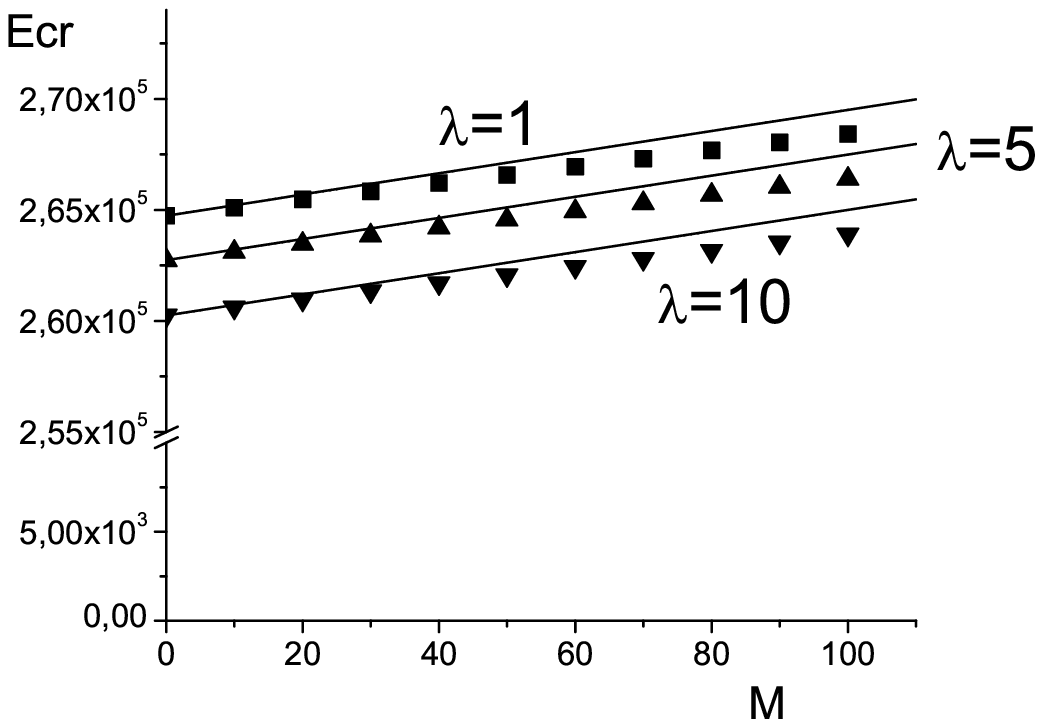}
\includegraphics[height=5.5cm]{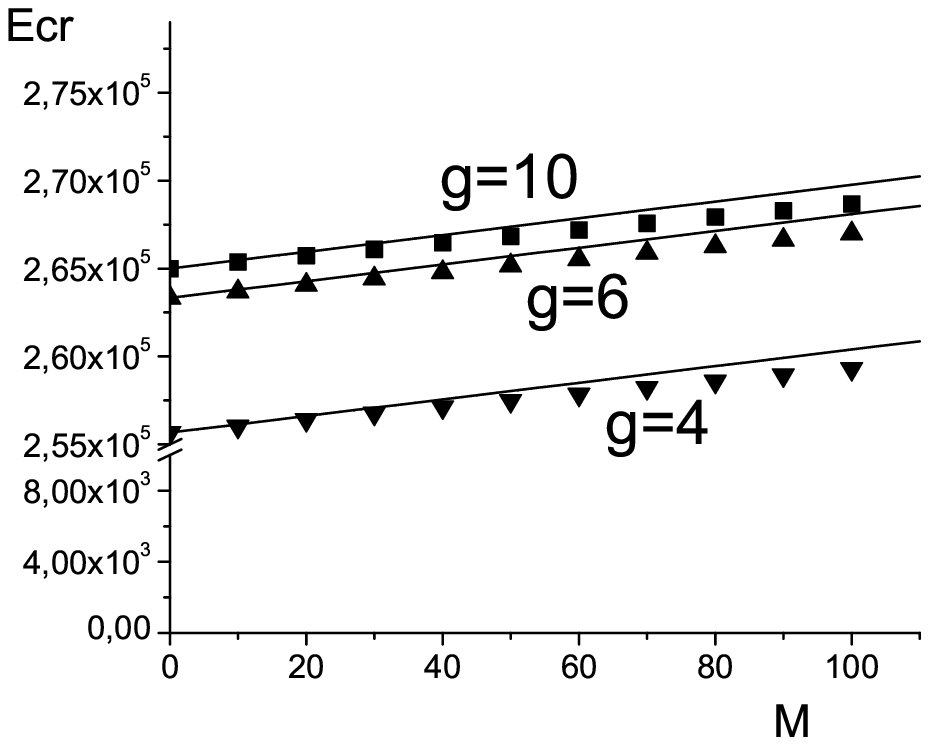}
\caption{Relation between critical density of energy and
centrifugal constant M at various values of self-coupling
constants of scalar field($\lambda$) and non-abelian gauge
fields($g$).}\label{Ecrl}
\end{figure}

Using well known technique based on Toda criterion of local
instability \cite{Salasnich} one can obtain dependence of critical
density of energy of order to chaos transition (minus vacuum density
of energy which is non-zero) on centrifugal constant $M$ and
coupling constants of Yang-Mills and Higgs fields. Numerical
calculations allow us to build two required plots. First plot
(figure~\ref{Ecrl}) demonstrates dependence  of critical density of
energy on value $M $ at various values of a self-coupling constant
of scalar field $ \lambda $(other constants have following values
$\mu=100,g=10$). Second plot (figure~\ref{Ecrl}) demonstrates
dependence on value of another self-coupling constant
$g$($\mu=100,\lambda=0.495$).  Points on this plots are in
accordance to obtaining numerical results.

From obtained plots we can make next conclusions. First, it is
clarified that taking into account centrifugal term ${M^2}/{r^2}$
increases the region of regular dynamics of Yang-Mills and Higgs
fields system at low densities of energy. And second, dependence
of value of critical density of energy on value M is approximate
linear at any prescribed values of self-coupling constants of
Yang-Mills and Higgs fields. And third, it is seen that
self-coupling constant $g$ more intensively then $ \lambda $
influences on dynamics of considered system. It is adjust with
results obtaining in work\cite {Previous}.

Further, basing on Toda criterion of local instability
\cite{Salasnich} and the following prerequisite $U\left( <\rho>
\right)'' = ~ 0$ \cite {Previous} we have found analytically the
approximate relation for critical density of energy in the case $M
\neq 0$

\begin{equation}\label{EM}
 E_{cr} = \frac{1}{\sqrt{2}} \mu \exp{\left( \frac{1}{2}
\alpha_w - \frac{1}{2} \frac{\lambda}{g^4}
  \beta_w - \frac{1}{3}\right)} M + E_c .
\end{equation}
Were we used the following denotations \cite {Previous}

\begin{eqnarray}\label{a}
\alpha_w = \frac{2 \ln{\cos{\theta_w}}}{1 + 2 \cos^{4}{\theta_w}},
\quad \quad \quad   \beta_w = \frac{32 \pi^2
\cos^{4}{\theta_w}}{9(1 + 2 \cos^{4}{\theta_w})}\label{b}
\end{eqnarray}
and value $E_c$  describes critical density of energy in case of
$M=0$\cite {Previous}

\begin{equation}\label{E}
 E_c = \frac{3 \mu^4 }{64 \pi^2} \exp{\left( 2 \alpha_w - \frac{2\lambda}{g^4}
  \beta_w \right)} \left( 1 +\frac{1}{2 \cos ^{4}{\theta _{w}} } \right)
  \left( 1 -\frac{7}{3}{ } e^{-\frac43} \right) .
\end{equation}

To check-up relation (\ref{EM}) we have made a comparison between
results obtaining with mentioned relation and acquired
numerically(figure~\ref{Ecrl}). Points on this plots accordance to
numerical results and lines to analytical ones. From mentioned plots
it is seen that errors are proportional to the value $M$ and
obtained relation (\ref{EM}) describes closely numerical results in
a region of small value $M$.

In conclusion, we have considered $SU(2)\bigotimes U(1)$ gauge field
theory describing electroweak interactions. We have demonstrated
that centrifugal term increases the region of regular dynamics of
Yang-Mills and Higgs fields system at low densities of energy. It is
necessary to note that mentioned increase of the region of regular
dynamics has linear dependance on the value M. The approximate
expression for critical density of energy(\ref{EM}) was also
analytically found. Obtained results show us Higgs field influences
on the stability of Yang-Mills fields dynamics.

\end{document}